\documentclass[12pt,a4paper,final]{iopart}
\usepackage{iopams} 
\usepackage{graphicx}
\usepackage{hyperref}
\usepackage{cleveref}
\begin{document}
\title[Quasi local entities]{Aspects of Quasi-local energy for gravity coupled to gauge fields}
\author{Puskar Mondal$^{2}$ \footnote{e-mail: puskar$\_$mondal@fas.harvard.edu},~Shing-Tung Yau$^{1,2}$ \footnote{e-mail: yau@math.harvard.edu }}

\begin{abstract}
We study the aspects of quasi-local energy associated with a $2-$surface $\Sigma$ bounding a space-like domain $\Omega$ of a physical $3+1$ dimensional spacetime in the regime of gravity coupled to a gauge field. The Wang-Yau quasi-local energy together with an additional term arising due to the coupling of gravity to a gauge field constitutes the total energy ($\mathcal{QLE}$) contained within the membrane $\Sigma=\partial\Omega$. We specialize in the Kerr-Newman family of spacetimes which contains a U(1) gauge field coupled to gravity and an outer horizon. Through explicit calculations, we show that the total energy satisfies a weaker version of a Bekenstein type inequality $\mathcal{QLE}> \frac{Q^{2}}{2R}$ for large spherical membranes, $Q$ is the charge and $R$ is the radius of the membrane. Turning off the angular momentum (Reissner Nordstr\"om) yields $\mathcal{QLE}> \frac{Q^{2}}{2R}$ for all constant radii membranes containing the horizon and in such case the charge factor appearing in the right-hand side exactly equals to that of Bekenstein's inequality. Moreover, we show that the total quasi-local energy monotonically decays from $2M_{irr}+V_{Q}$ ($M_{irr}$ is the irreducible mass, $V_{Q}$ is the electric potential energy) at the outer horizon to $M$ ($M$ is the ADM mass) at the space-like infinity under the assumption of a small angular momentum of the black hole.
\end{abstract}

\section{Introduction}
The notion of a local mass density of pure gravity is non-existent due to the equivalence principle. For an isolated self-gravitating system where the spacetime is asymptotically flat, one defines the notion of mass as a flux integral over a space-like topological 2-sphere located at infinity. For such a system, this so-called ADM mass satisfies the desired positivity property \cite{schoen1979proof, schoen1981proof, witten1981new}. One also defines a notion of mass at the null-infinity \cite{bondi1962gravitational,schoen1982proof} as well. On the other hand, a notion of mass in-between scales is of extreme importance due to the fact that most physical models extend over a finite region. For example, if one were to study the kinematics of a black hole on a curved background, the first entity that one would require is the \textit{mass} associated with a topological $2-$ sphere that encloses the black hole. There are several problems in classical general relativity that require a notion of quasi-local mass as well. Penrose's singularity theorem \cite{penrose1965gravitational} essentially entails the study of the dynamics of space-like $2-$surfaces foliating the future null cone of an arbitrary point in the spacetime. In other words, the formation of black-hole is hinted at by the congruence of the future null geodesics measured by the trace of the null second fundamental form of topological $2-$spheres foliating the outgoing null-hypersurfaces generated by such null geodesics(later \cite{christodoulou2012formation} proved the dynamical formation of such trapped surfaces from initial data that did not contain the `trapped' characteristics in case of pure vacuum gravity). In such a case, one would want to study the evolution of the gravitational energy that is contained within such topological $2-$spheres and understand if such an evolution exhibits any special characteristics that hint towards a possible singularity formation. In order to even study such an evolution problem, one would need an expression of the energy contained within the surface of interest necessitating a formulation of \textit{quasi-}local mass/energy. In addition, a formulation (let alone proof) of the hoop conjecture \cite{thorne1972nonspherical, gibbons2009birkhoff} of general relativity requires a notion of quasi-local mass since this conjecture essentially deals with the question of the implosion of an object to a limit where a circular hoop of circumference $2\pi r_{s}$ ($r_{s}$ being the Schwarzschild radius) can be placed around it. Such a limit corresponds to a point of no return or eventual black hole formation similar to the formation of a trapped surface. One requires a notion of mass of the  object under study that can be formulated as a quasi-local mass of a topological $2-$sphere bounding the object to address such conjecture. In addition, the study of black hole collision and merging also requires an appropriate notion of quasi-local mass. 

Motivated by the aforementioned physically relevant questions, several authors defined different notions of quasi-local mass over the years. A necessary property of such a quasi-local mass would be that it should be positive for a space-like $2-$surface embedded in a curved spacetime and vanishes identically for any such $2-$surface in the Minkowski space. In addition, it should encode the information about pure gravitational energy (Weyl curvature effect) as well as the stress-energy tensor of any source fields present in the spacetime. Based on a Hamilton-Jacobi analysis, Brown-York \cite{brown1992quasilocal, brown1993quasilocal} and Li-Yau \cite{liu2003positivity,liu2006positivity} defined a quasi-local mass by isometrically embedding the $2-$surface into the reference Euclidean $3-$ space and comparing the extrinsic geometry (the formulation relied on the embedding theorem of Pogorelov \cite{pogorelov1952regularity} i.e., the topological $2-$sphere needed to posses everywhere non-negative sectional curvature). However, \cite{murchadha2004comment} discovered surfaces in Minkowski space that do have strictly positive Brown-York and Li-Yau mass. It appeared that these formulations were lacking a prescription of momentum information. This led Wang and Yau \cite{Wangyau, wang2009quasilocal} to define the most consistent notion (till today) of the quasi-local mass associated with a space-like topological $2-$surface. It involves isometric embedding of the topological $2-$sphere bounding a space-like domain in the physical spacetime satisfying dominant energy condition (energy can not flow into a past light cone of an arbitrary point in spacetime; essentially finite propagation speed) into the Minkowski spacetime instead of Euclidean $3-$space. This formulation relies on a weaker condition on the sectional curvature of the $2-$surface of interest and solvability of Jang's equation \cite{jang1978positivity} with prescribed Dirichlet boundary data. The Wang-Yau quasi-local mass is then defined as the infimum of the Wang-Yau quasi-local energy among all physical observers that is found by solving an optimal isometric embedding equation. This Wang-Yau quasi-local mass possesses several good properties that are desired on a physical ground. This mass is strictly positive for $2-$surfaces bounding a space-like domain in a curved spacetime that satisfies the dominant energy condition and it identically vanishes for any such $2-$surface in Minkowski spacetime. In addition, it coincides with the ADM mass at space-like infinity \cite{wang2010limit} and Bondi-mass at null infinity \cite{chen2011evaluating}, and reproduces the time component of the Bel-Robinson tensor (a pure gravitational entity) together with the matter stress-energy at the small sphere limit, that is when the $2-$sphere of interest is evolved by the flow of its null geodesic generators and the vertex of the associated null cone is approached \cite{chen2018evaluating}. In addition, explicit conservation laws were also discovered at the asymptotic infinity \cite{chen2015conserved}.         

These physically desired features led to the belief that Wang-Yau quasi-local mass may be the one to consider as an appropriate notion of a quasi-local mass modulo few technicalities such as the mean curvature vector of the $2-$surface of interest is restricted to be space-like \cite{Wangyau, wang2009quasilocal}. Therefore, it is important to apply the Wang-Yau formalism to physically interesting spacetimes, explicitly compute the mass of a space-like domain bounded by a topological $2-$surface, and understand its properties. In this article, we are interested in spacetimes where an additional gauge field is coupled to gravity. The gauge fields are particularly interesting since their evolution is not free due to gauge invariance and therefore constraints must be solved on each Cauchy hypersurface (spatial hypersurface that is met exactly once by every inextendible causal curve) mush like the pure gravity problem itself. Due to these constraints, additional boundary terms appear in the expression of quasi-local energy through Hamilton-Jacobi analysis that is not controlled by the Wang-Yau quasi-local energy expression (unlike sources with non-gauge degrees of freedom where Wang-Yau quasi-local energy is sufficient to control their energy content in addition to pure gravitational energy as well). This additional contribution arises as a consequence of non-vanishing \textit{charge} associated with the gauge field. Therefore it turns out to be extremely important in the context of charged black-holes (electromagnetic or Yang-Mills) since we expect there must be a charge-mass inequality in order for the interior singularity to be hidden or in technical terms, the black-holes should be of sub-extremal type since exposing the interior singularity to a causal observer located in the domain of outer communication signals a pathological breakdown of the classical general relativity. Since we consider the notion of mass to be Wang-Yau quasi-local mass, this translates to obtaining a suitable inequality relating the charge of the gauge field and the W-Y quasi-local mass of membranes enclosing the black hole.       

In addition, one of the other main motivation of the current study is the Bekenstein's inequality relating total energy of a relativistic object bounded by a membrane, its charge, and the angular momentum. Based on physical arguments, Bekenstein proposed an upper bound of the entropy of the object which takes the following form in the natural unit $\hbar=c=\kappa=1$ \cite{bekenstein2020universal} 
\begin{eqnarray}
\mathcal{S}\leq \sqrt{\mathcal{E}^{2}R^{2}-\mathcal{J}^{2}}-\frac{Q^{2}}{2},
\end{eqnarray}
where $\mathcal{S},\mathcal{E},\mathcal{J},Q$, and $R$ are the objects entropy, total energy, angular momentum, charge, and size (the radius of smallest sphere containing it), respectively. This inequality proves to be difficult to establish in a rigorous way. Nevertheless, if one simply assumes the non-negativity of the entropy, then a weaker version of the above inequality reads 
\begin{eqnarray}
\mathcal{E}^{2}\geq \frac{\mathcal{J}^{2}}{R^{2}}+\frac{Q^{4}}{4R^{2}}
\end{eqnarray}
and for an object with non-negative angular momentum
\begin{eqnarray}
\mathcal{E}\geq \frac{Q^{2}}{2R}.
\end{eqnarray}
As a first step of the proof of this inequality, one requires a physically reasonable definition of the energy contained within a region of non-zero charge bounded by a membrane in the fully general relativistic setting. A natural choice would be to consider the Wang-Yau quasi-local energy and try to establish the previous inequality. We note that recently \cite{alaee2019geometric} studied this inequality for several definitions of the quasi-local energy.   

The structure of the article is as follows. Starting with the classical Hilbert action of the Einstein-Yang-Mills system (including a $U(1)$ Yang-Mills as well), we derive the boundary contribution to the quasi-local energy that arises solely due to the gauge fields through an ADM Hamilton-Jacobi analysis. Later, we specialize in the Kerr-Newman spacetime and explicitly compute the boundary contribution. Simultaneously, we compute the Wang-Yau energy functional and establish the necessary condition for it to be well behaved. We later show that the trivial data solves the optimal isometric embedding equation and the associated mass is positive by construction if the angular momentum of the black hole is not too large. This yields an inequality between the charge of the black hole and the quasi-local energy of membranes enclosing it. This total quasi-local energy is shown to exhibit monotonic decay in the radially outward direction from the outer horizon under the assumption of a small angular momentum.  A few physical conclusions are drawn on the basis of our results and a few additional problems are discussed that are to be handled in the potential future.        

\section{Quasi-local energy expression including a gauge field}
In this section, we derive the contribution of the gauge field to the expression of the quasi-local mass. We consider a `1+3' dimensional $C^{\infty}$ globally hyperbolic spacetime $\mathbb{R}\times M$, where $M$ is diffeomorphic to a Cauchy hypersurface, equipped with a Lorentzian metric $\hat{g}$. Since this globally hyperbolic spacetime is foliated by the space-like submanifolds $M_{t}$ ($t$ is a \textit{time} function which is well defined for a globally hyperbolic spacetime), we may use a \textit{lapse} function $N$ and a $M-$parallel \textit{shift} vector field $Y$ to represent $\hat{g}$ in the following ADM form \cite{arnowitt2008republication} 
\begin{eqnarray}
\label{eq:metriclabel}
\hat{g}:=-N^{2}dt\otimes dt+g_{ij}(dx^{i}+Y^{i}dt)\otimes (dx^{j}+Y^{j}dt),
\end{eqnarray}
where $g$ is the Riemannian metric induced on $M$ by the embedding $i:M\hookrightarrow \mathbb{R}\times M$. The second fundamental form $K$ measuring the extrinsic geometry of $M$ in $\mathbb{R}\times M$ is defined by $K(X,Z):=-\hat{g}(\nabla[\hat{g}]_{X}\mathbf{n},Z),$ for all $X,Z$ being the sections of $TM$. Here $\mathbf{n}$ is the $t=$constant hypersurface orthogonal unit time-like vector field defined as $\mathbf{n}:\frac{1}{N}(\partial_{t}-Y)$.

To formulate the Yang-Mills theory over the spacetime $\mathbb{R}\times M$, we first choose a compact semisimple Lie group $G$. If a section of the principle $G-$bundle defined over $\mathbb{R}\times M$ is chosen and the connection is pulled back to the base manifold, then it yields a $1-$form field on the base which takes values in the Lie algebra $\mathfrak{g}$ of $G$. Let us consider the dimension of the group $G$ to be $dim_{G}$ and since $\mathfrak{g}:=T_{e}G$, it has a natural vector space structure. Assume that the vector space $\mathfrak{g}$ has a basis $\{\chi_{A}\}_{A=1}^{dim_{G}}$ given by a set of $k\times k$ real-valued matrices ($k$ being the dimension of the representation $V$ of the Lie algebra $\mathfrak{g}$). The connection $1-$form field is then defined to be 
\begin{eqnarray}
\hat{A}:=\hat{A}^{A}_{\mu}\chi_{A}dx^{\mu}=\hat{A}^{A}_{\mu}(\chi_{A})^{a}_{b}dx^{\mu}=\hat{A}^{a}~_{b\mu}dx^{\mu},~a,b=1,2,3,...,k.
\end{eqnarray}
From now on by the connection 1-form field $\hat{A}_{\mu}$, we will always mean $\hat{A}^{a}~_{b\mu}$. In the current setting $\hat{A}\in \Omega^{1}(\mathbb{R}\times M;End(V))$, where $End(V)$ denotes the space of endomorphisms of the vector space $V$. The curvature of this connection is defined to be the Yang-Mills field $F\in \Omega^{2}(\mathbb{R}\times M; End(V))$
\begin{eqnarray}
\hat{F}^{a}~_{b\mu\nu}:=\partial_{\mu}\hat{A}^{a}~_{b\nu}-\partial_{\nu}\hat{A}^{a}~_{b\mu}+[\hat{A},\hat{A}]^{a}~_{b\mu\nu},
\end{eqnarray}
where the bracket is defined on the Lie algebra $\mathfrak{g}$ and given by commutator of matrices under multiplication. The Yang-Mills coupling constant is set to unity. Since $G$ is compact, it admits a positive definite adjoint invariant metric on $\mathfrak{g}$. We choose a basis of $\mathfrak{g}$ such that this adjoint invariant metric takes the Cartesian form $\delta_{AB}$ and work with representations for which the bases satisfy 
\begin{eqnarray}
-\tr(\chi_{A}\chi_{B})=(\chi_{A})^{a}_{b}(\chi_{A})^{b}_{a}=\delta_{AB}.
\end{eqnarray}     
One may for convenience decompose the $1+3$ $\mathfrak{g}$-valued connection $1-$form field $A^{a}~_{b\mu}$ into its component parallel and perpendicular to $M$ as follows 
\begin{eqnarray}
\hat{A}^{a}~_{b}=A^{a}~_{b}-\hat{g}(A^{a}~_{b},\mathbf{n})\mathbf{n}
\end{eqnarray}
where $A^{a}~_{b}$ is a $\mathfrak{g}$ valued $1-$form field parallel to the spatial manifold $M$ i.e., $A\in \Omega^{1}(M;End(V))$. Importantly note that $\hat{A}^{a}~_{bi}=A^{a}~_{bi}$ but $\hat{A}^{a}~_{b}^{i}\neq A^{a}~_{b}^{i}$ unless the shift vector field $Y$ vanishes. Similarly, we decompose the Yang-Mills field strength $\hat{F}$ as follows 
\begin{eqnarray}
\hat{F}^{a}~_{b}=\frac{1}{\mu_{g}}(\mathcal{E}^{a}~_{b}\otimes \mathbf{n}-\mathbf{n}\otimes\mathcal{E}^{a}~_{b})+F^{a}~_{b},
\end{eqnarray}
where $\mathcal{E}\in \Omega^{1}(M;End(V))$ is the electric field, $F\in \Omega^{2}(M;End(V))$ is related to the magnetic field, and $\mu_{g}:=\sqrt{\det(g_{ij})}$. Once again note $\hat{F}^{a}~_{bij}=F^{a}~_{bij}$ but $\hat{F}^{a}~_{b}~^{ij}\neq F^{a}~_{b}~^{ij}$. The Einstein-Hilbert action for a gauge-gravity system in the natural unit $G=1=\wp$ ($\wp$ is the Yang-Mills coupling constant) may be written as follows \cite{hawking1996gravitational}
\begin{eqnarray}
\label{eq:hilbert}
8\pi\mathcal{S}&:=&\frac{1}{2}\int_{I\times M} \mathcal{R}(\hat{g})\mu_{\hat{g}}-\frac{1}{4}\int_{I\times M}\hat{F}^{a}~_{b\mu\nu}F^{b}~_{a\alpha\beta}\hat{g}^{\alpha\mu}\hat{g}^{\beta\nu}\mu_{\hat{g}}\\\nonumber 
&&+\int_{M_{t}-M_{t_{0}}}\tr_{g}K\mu_{\hat{g}}|_{M},
\end{eqnarray}
where $I\subset \mathbb{R}$ denotes a closed interval on the real line i.e., $I:=[t_{0},t]$, $\mu_{\hat{g}}:=\sqrt{|\det(\hat{g}_{\mu\nu})|}$, $K\tr_{g}$ is the mean extrinsic curvature of the constant time hypersurface $M$, and $\mu_{\hat{g}}|_{M}$ is the volume form induced on $M$. $M_{t}-M_{t_{0}}$ denotes the difference of the integrals over hypersurfaces $t=t$ and $t=t_{0}$. Through the ADM decomposition (\ref{eq:metriclabel}), one may introduce the canonical pairs ($g_{ij},\pi^{ij}:=-\mu_{g}(K^{ij}-\tr_{g}K g^{ij})$) for the gravity and $(A^{a}~_{bi},\mathcal{E}^{a}~_{b}~^{i}:=\frac{\mu_{g}}{N}g^{ik}(\hat{F}^{a}~_{b0k}-\hat{F}^{a}~_{bjk}Y^{j}))$ for the Yang-Mills theory. Now, using the same ADM formalism, we may obtain an expression for the quasi-local energy associated with a $2-$surface bounding a domain in $M$ for this gauge-gravity coupled system through the Hamilton-Jacobi analysis. The following lemma describes the result.

\textbf{Lemma 1:} \textit{Let $\Sigma$ be a $2-$surface bounding a domain $\Omega$ in $M$ i.e., $\partial \Omega=\Sigma$ and $\nu$ be its space-like outward unit normal vector.  Also assume that the Einstein-Hilbert action (\ref{eq:hilbert}) for a gauge-gravity coupled system is defined on $[t_{0},t]\times \Omega$. Then the quasi-local Hamiltonian defined by $\mathcal{H}_{ql}:=-\partial_{t}\mathcal{S}$ for this Einstein-Yang-Mills system with a compact semi-simple gauge group verifies the following expression 
\begin{eqnarray}
\mathcal{H}_{ql}:=-\frac{\partial \mathcal{S}}{\partial t}=-\frac{1}{8\pi}\int_{\Sigma_{t}}(kN-\frac{\pi^{ij}}{\mu_{\Sigma}}\nu_{i}Y_{j}-\frac{\mathcal{E}^{a}~_{b}~^{i}}{\mu_{\Sigma}}A^{b}~_{a0}\nu_{i})\mu_{\Sigma}d^{2}x,
\end{eqnarray}
where $k=\tr_{g}K$, and $\mu_{\Sigma}$ is the induced volume form on $\Sigma$. 
}

\textbf{Proof:} Let us start with the Einstein-Hilbert action $\mathcal{S}:=\frac{1}{2}\int_{[t_{0},t]\times \Omega} \mathcal{R}(\hat{g})\mu_{\hat{g}}-\frac{1}{4}\int_{[t_{0},t]\times \Omega}\hat{F}^{a}~_{b\mu\nu}F^{b}~_{a\alpha\beta}\hat{g}^{\alpha\mu}\hat{g}^{\beta\nu}\mu_{\hat{g}}+\int_{\Omega_{t}-\Omega_{t_{0}}}\tr_{g}K\mu_{\hat{g}}|_{\Omega}$ and reduce it utilizing the ADM splitting (\ref{eq:metriclabel}). Straightforward calculations yield 
\begin{eqnarray}
\mathcal{R}(\hat{g})=R(g)+K_{ij}K^{ij}-(\tr_{g}K)^{2}-2(\nabla_{\mu}(n^{\nu}\nabla_{\nu}n^{\mu})-\nabla_{\nu}(n^{\nu}\nabla_{\mu}n^{\mu}))
\end{eqnarray}
and 
\begin{eqnarray}
\frac{1}{4}\hat{F}^{a}~_{b\mu\nu}F^{b}~_{a\alpha\beta}&=&-\frac{1}{2N^{2}}g^{ik}(\hat{F}^{a}~_{b0i}-F^{a}~_{bji}X^{j}) (\hat{F}^{b}~_{a0k}-F^{b}~_{ajk}X^{j})\\\nonumber &&+\frac{1}{4}g^{ik}g^{jm}F^{a}~_{bij} F^{b}~_{akm},
\end{eqnarray}
where we have used the fact that $\hat{F}^{a}~_{bij}=F^{a}~_{bij}$.
Now recall the definitions of the gravitational and Yang-Mills momenta 
\begin{eqnarray}
\pi^{ij}:=-\mu_{g}(K^{ij}-\tr_{g}Kg^{ij}),~\mathcal{E}^{a}_{b}~^{i}:=\frac{\mu_{g}}{N}g^{ik}(\hat{F}^{a}~_{b0k}-F^{a}~_{bjk}Y^{j}),
\end{eqnarray}
substitution of which yields the following expression for the action $\mathcal{S}(t)$
\begin{eqnarray}
8\pi\mathcal{S}(t)&=&\frac{1}{2}\int_{[t_{0},t]}\int_{\Omega}\left(\pi^{ij}\partial_{t}g_{ij}\nonumber-N\mu_{g}(|K|^{2}_{g}-(\tr_{g}K)^{2}-R(g))-2\pi^{ij}\nabla_{i}Y_{j}\right)d^{3}xdt\\\nonumber
&&+\int_{I}\int_{\Omega}\left(\mathcal{E}^{a}~_{b}~^{i}\partial_{t}A^{b}~_{ai}-\mathcal{E}^{a}~_{bi}F^{b}~_{aji}Y^{j}+\mathcal{E}^{a}~_{b}~^{i}[A_{0},A_{i}]^{b}~_{a}-\frac{N}{2\mu_{g}}\mathcal{E}^{a}~_{b}~^{i}\mathcal{E}^{b}~_{ai}\right.\\\nonumber
&&\left.-\frac{1}{4}F^{a}~_{bij}F^{b}~_{a}~^{ij}N\mu_{g}-\mathcal{E}^{a}~_{b}~^{i}\partial_{i}A^{b}~_{a0}\right)d^{3}xdt\\\nonumber
&&+\int_{[t_{0},t]}\int_{\Omega}kN\mu_{\Sigma}d^{2}xdt.
\end{eqnarray}
Here we have used the definition of the second fundamental form $K_{ij}=\hat{g}(\nabla[\hat{g}]_{\partial_{i}}n,\partial_{j})=-\frac{1}{2N}(\partial_{t}g_{ij}-L_{Y}g_{ij})$ and integration by parts of $(\nabla_{\mu}(n^{\nu}\nabla_{\nu}n^{\mu})-\nabla_{\nu}(n^{\nu}\nabla_{\mu}n^{\mu}))N\mu_{g}$ to obtain the last boundary term since the first term cancels out by $\int_{\Omega_{t}-\Omega_{t_{0}}}\tr_{g}K\mu_{\hat{g}}|_{\Omega}$. Now notice the following calculations 
\begin{eqnarray}
\int_{\Omega}\pi^{ij}\nabla_{i}Y_{j}d^{3}x=\int_{\Omega}\nabla_{i}(\pi^{ij}Y_{j})d^{3}x-\int_{\Omega}Y_{j}\nabla_{i}\pi^{ij}d^{3}x,
\end{eqnarray}
where we may integrate the first term since $\pi$ is a density and therefore $\nabla_{i}(\pi^{ij}Y_{j})=\partial_{i}(\pi^{ij}Y_{j})$ yielding
\begin{eqnarray}
\int_{\Omega}\pi^{ij}\nabla_{i}Y_{j}d^{3}x=\int_{\Sigma}\pi^{ij}Y_{j}\nu_{i}d^{2}x-\int_{\Omega}Y_{j}\nabla_{i}\pi^{ij}d^{3}x.
\end{eqnarray}
In an exact similar way we may reduce the similar term of the Yang-Mills sector 
\begin{eqnarray}
\int_{\Omega}\mathcal{E}^{a}~_{b}^{i}\partial_{i}A^{b}~_{a0}d^{3}x&=&\int_{\Omega}\partial_{i}(\mathcal{E}^{a}~_{b}^{i}A^{b}~_{a0})d^{3}x-\int_{\Omega}\partial_{i}\mathcal{E}^{a}~_{b}^{i} A^{b}~_{a0}d^{3}x\\\nonumber 
&=&\int_{\Omega}\mathcal{E}^{a}~_{b}~^{i}A^{b}~_{a0}\mathcal{\nu}_{i}d^{2}x-\int_{\Omega}\partial_{i}\mathcal{E}^{a}~_{b}^{i} A^{b}~_{a0}d^{3}x.
\end{eqnarray}
Notice that the terms $\int_{\Omega}Y_{j}\nabla_{i}\pi^{ij}d^{3}x$ and $\int_{\Omega}\partial_{i}\mathcal{E}^{a}~_{b}^{i} A^{b}~_{a0}d^{3}x$ give rise to the momentum constraint and the Gauss law constraint, respectively. Assembling all the terms together, we obtain the final expression for the action functional 
\begin{eqnarray}
8\pi\mathcal{S}(t)=\frac{1}{2}\int_{[t_{0},t]}\int_{\Omega}\left(\pi^{ij}\partial_{t}g_{ij}\nonumber-N\mu_{g}(|K|^{2}_{g}-k^{2}-\mathcal{R}(g)+\frac{1}{\mu^{2}_{g}}\mathcal{E}^{a}~_{b}~^{i}\mathcal{E}^{b}~_{ai}\right.\\\nonumber\left.+\frac{1}{2}F^{a}~_{bij}F^{b}~_{a}~^{ij})+(2\nabla_{i}\pi^{i}_{j}-\mathcal{E}^{a}~_{b}^{i}F^{b}~_{aji})Y^{j}\right)d^{3}xdt\\\nonumber+\int_{[t_{0},t]}\underbrace{\int_{\Sigma}(kN-\frac{\pi^{ij}}{\mu_{\Sigma}}\nu_{i}Y_{j})\mu_{\Sigma} d^{2}x}_{gravitational~energy~modulo~a~reference~term}dt\\\nonumber 
+\int_{[t_{0},t]}\int_{\Omega}\left(\mathcal{E}^{a}~_{b}~^{i}(\partial_{t}A^{b}~_{ai}+[A_{0},A_{i}]^{b}~_{a})+\partial_{i}\mathcal{E}^{a}~_{b}~^{i}A^{b}~_{a0}\right)d^{3}xdt\\\nonumber 
-\int_{[t_{0},t]}\underbrace{\int_{\Sigma}\mathcal{E}^{a}~_{b}~^{i}A^{b}~_{a0}\mathcal{\nu}_{i}d^{2}x}_{surface~energy~from~gauge~field}dt.
\end{eqnarray}
Therefore using the Hamilton-Jacobi equation $\partial_{t}\mathcal{S}+\mathcal{H}_{q.l}=0$, we obtain
\begin{eqnarray}
\mathcal{H}_{ql}=-\frac{1}{8\pi}\int_{\Sigma}(kN-\frac{\pi^{ij}}{\mu_{\Sigma}}\nu_{i}Y_{j}-\frac{\mathcal{E}^{a}~_{b}~^{i}}{\mu_{\Sigma}}A^{b}~_{a0}\nu_{i})\mu_{\Sigma}d^{2}x.
\end{eqnarray}
This concludes the proof of the lemma. ~~~~~~~~~~~~~~~~~~~~~~~~~~~~~~~~~~~~~~$\Box$

Wang and Yau \cite{Wangyau, wang2009quasilocal} dealt with the pure gravitational and non-gauge part of the quasi-local energy by subtracting a well-defined reference contribution (by isometrically embedding $\Sigma$ into Minkowski space).  Therefore, we will use their expression to handle the first term $-\int_{\Sigma}(kN-\frac{\pi^{ij}}{\mu_{\Sigma}}\nu_{i}Y_{j})\mu_{\Sigma}d^{2}x$ while computing the quasi-local energy for a given spacetime. They have shown with a suitable choice of the lapse function $N$ and the shift vector field $Y$ that one can define a notion of quasi-local mass which satisfies several properties that are desired on the physical ground. Let us now describe the Wang-Yau quasi-local mass which we wish to evaluate. Let us assume that the mean curvature vector $\mathbf{H}$ of $\Sigma$ is space-like. Let $\mathbf{J}$ be the reflection of $\mathbf{H}$ through the future outgoing light cone in the normal bundle of $\Sigma$. The data that Wang and Yau use to define the quasi-local energy is the triple ($\sigma,|\mathbf{H}|_{\hat{g}},\alpha_{\mathbf{H}}$) on $\Sigma$, where $\sigma$ is the induced metric on $\Sigma$ by the Lorentzian metric $\hat{g}$ or $\mathbb{R}\times M$, $|\mathbf{H}|_{\hat{g}}$ is the Lorentzian norm of $\mathbf{H}$, and $\alpha_{\mathbf{H}}$ is the connection $1-$form of the normal bundle with respect to the mean curvature vector $\mathbf{H}$ and is defined as follows 
\begin{eqnarray}
\label{eq:connection}
\alpha_{\mathbf{H}}(X):=\hat{g}(\nabla[\hat{g}]_{X}\frac{\mathbf{J}}{|\mathbf{H}|},\frac{\mathbf{H}}{|\mathbf{H}|}).
\end{eqnarray}
Choose a basis pair ($e_{3},e_{4}$) for the normal bundle of $\Sigma$ in the spacetime that satisfy $\hat{g}(e_{3},e_{3})=1,\hat{g}(e_{4},e_{4})=-1$, and $\hat{g}(e_{3},e_{4})=0$. Now embed the $2-$surface $\Sigma$ isometrically into the Minkowski space with its usual metric $\eta$ i.e., the embedding map $X: x^{a}\mapsto X^{\mu}(x^{a})$ satisfies $\sigma(\frac{\partial}{\partial x^{a}},\frac{\partial}{\partial x^{b}})=\langle \frac{\partial X}{\partial x^{a}},\frac{\partial X}{\partial x^{b}}\rangle_{\eta}$, where $\{x^{a}\}_{a=1}^{2}$ are the coordinates on $\Sigma$. Now identify a basis pair ($e_{30},e_{40}$) in the normal bundle of $X(\Sigma)$ in the Minkowski space that satisfy the exact similar property as $(e_{3},e_{4})$. In addition the time-like unit vector $e_{4}$ is chosen to be future directed i.e., $\hat{g}(e_{4},\partial_{t})<0$. Let $\tau:=-\langle X,\partial_{t}\rangle_{\eta}$, a function on $\Sigma$ be the time function of the embedding $X$. 
The Wang-Yau quasi-local energy is defined as follows 
\begin{eqnarray}
\mathcal{QLE}_{gravity}=\frac{1}{8\pi}\underbrace{\int_{\Sigma_{t}}\left(-\sqrt{1+|\nabla\tau|^{2}_{\sigma}}\langle \mathbf{H}_{0},e_{30}\rangle-\langle\nabla[\eta]_{\nabla\tau}e_{30},e_{40}\rangle\right)\mu_{\Sigma}}_{I:=Contribution~from~the~Minkowski~space}\\\nonumber 
-\frac{1}{8\pi}\underbrace{\int_{\Sigma_{t}}\left(-\sqrt{1+|\nabla\tau|^{2}_{\sigma}}\langle \mathbf{H},e_{3}\rangle-\langle\nabla[\hat{g}]_{\nabla\tau}e_{3},e_{4}\rangle\right)\mu_{\Sigma}}_{II:=contribution~from~the~physical~spacetime}.
\end{eqnarray}
However, there is still a gauge redundancy due to the boost transformations in the normal bundle of $\Sigma$. In other words, one is left with the freedom of choosing $e_{3}$ and $e_{4}$ since one may apply a hyperbolic rotation (boost) to yield another pair $(\hat{e}_{3},\hat{e}_{4})$. Wang and Yau \cite{Wangyau, wang2009quasilocal} considers the following minimization procedure to get rid of this extra gauge freedom. Choose a fixed basis $(\hat{e}_{3},\hat{e}_{4})$ of the fibres of the normal bundle of $\Sigma$ such that the space-like mean curvature vector $\mathbf{H}$ is expressible as
\begin{eqnarray}
\label{eq:property}
\mathbf{H}=-|\mathbf{H}|_{\hat{g}}\hat{e}_{3},
\end{eqnarray}
where $\langle\hat{e}_{3},\hat{e}_{3}\rangle_{\hat{g}}=1$ and $\langle \hat{e}_{4},\hat{e}_{4}\rangle_{\hat{g}}=-1$. Write any general basis $(e_{3},e_{4})$ through a boost transformation in the normal bundle as follows 
\begin{eqnarray}
e_{3}=\cosh\psi \hat{e}_{3}-\sinh\psi \hat{e}_{4},\\ 
e_{4}=-\sinh\psi \hat{e}_{3}+\cosh\psi \hat{e}_{4} 
\end{eqnarray}
and substitute these expressions of $e_{3}$ and $e_{4}$ in the expression of $II$, and use the fact that $\hat{g}(\nabla[\hat{g}]_{e_{3}}\hat{e}_{4},e_{3})+\hat{g}(\nabla[\hat{g}]_{e_{4}}\hat{e}_{4},e_{4})=0$ (by construction) to yield
\begin{eqnarray}
II=\int_{\Sigma_{t}}\left(\sqrt{1+|\nabla\tau|^{2}_{\sigma}}|\mathbf{H}|_{\hat{g}}\cosh\psi-\alpha_{\hat{e}_{3}}(\nabla\tau)+\psi\Delta \tau\right)\mu_{\Sigma}.
\end{eqnarray}
This is a convex functional of $\psi$ and therefore is minimized for the following boost
\begin{eqnarray}
\psi=\sinh^{-1}(-\frac{\Delta \tau}{|\mathbf{H}|\sqrt{1+|\nabla\tau|^{2}_{\sigma}}}).
\end{eqnarray}
One repeats the same procedure for the surface with metric $\sigma$ embedded in the Minkowski space and write the Wang-Yau quasi-local energy as follows
\begin{eqnarray}
\mathcal{QLE}_{gravity}=\frac{1}{8\pi}\int_{\Sigma}\left(\sqrt{1+|\nabla\tau|^{2}_{\sigma}}|\mathbf{H}_{0}|_{\eta}\cosh\psi_{0}-\alpha_{\hat{e}_{30}}(\nabla\tau)\nonumber+\psi_{0}\Delta \tau\right)\mu_{\Sigma}\\-\frac{1}{8\pi}\int_{\Sigma}\left(\sqrt{1+|\nabla\tau|^{2}_{\sigma}}|\mathbf{H}|_{\hat{g}}\cosh\psi-\alpha_{\hat{e}_{3}}(\nabla\tau)+\psi\Delta \tau\right)\mu_{\Sigma}.
\end{eqnarray}
The quasi-local mass is defined to be the minimum of $\mathcal{QLE}_{gravity}$ in the space of the residual embedding function $\tau$. This is so because the isometric embedding of a $2-$surface into a $4$-dimensional manifold provides three constraints out of a total of four degrees of freedom. Therefore, the additional leftover degrees of freedom (in this case $\tau$) needs to be obtained by other physical means. In the current context, therefore, the mass is defined through a minimization procedure in the space of $\tau$ motivated by the definition of the rest mass which is the minimum in the space of observers. Through a variational argument, \cite{Wangyau} obtained the following fourth-order elliptic equation for $\tau$
\begin{eqnarray}
\label{eq:optimal}
-(\hat{H}\hat{\sigma}^{ab}-\hat{\sigma}^{ac}\hat{\sigma}^{bd}\hat{h}_{cd})\frac{\nabla[\sigma]_{a}\nabla_{b}\tau}{\sqrt{1+|\nabla\tau|^{2}_{\sigma}}}+\nabla[\sigma]^{a}(\frac{\nabla_{a}\tau \cosh\psi}{\sqrt{1+|\nabla\tau|^{2}_{\sigma}}}|\mathbf{H}|_{\hat{g}}\\\nonumber -\nabla_{a}\psi-(\alpha_{\hat{e}_{3}})_{a})=0,
\end{eqnarray}
where $\hat{h}$ is the second fundamental form of $\hat{\Sigma}$ while viewed as a surface in $\mathbb{R}^{3}$. From now on we will use the term mass and energy interchangeably if there is no confusion. The compatible condition that is required to guarantee the isometric embedding is the positivity of the Gauss curvature of $\hat{\Sigma}$ i.e., $K_{\Sigma}+(1+|\nabla\tau|^{2}_{\sigma})^{-1}\det(\nabla[\sigma]_{a}\nabla_{b}\tau)>0$, where $K_{\Sigma}$ is the Gauss curvature of $\Sigma$. 

Including the contribution from the pure gauge (Yang-Mills) part, we define the total quasi-local energy as follows
\begin{eqnarray}
\label{eq:gauge-gravity}
\mathcal{QLE}:=\mathcal{QLE}_{gravity}+\mathcal{QLE}_{gauge}\\\nonumber
=\frac{1}{8\pi}\int_{\Sigma}\left(\sqrt{1+|\nabla\tau|^{2}_{\sigma}}|\mathbf{H}_{0}|_{\eta}\cosh\psi_{0}-\alpha_{\hat{e}_{30}}(\nabla\tau)+\psi_{0}\Delta \tau\right)\mu_{\Sigma}\\\nonumber-\frac{1}{8\pi}\int_{\Sigma}\left(\sqrt{1+|\nabla\tau|^{2}_{\sigma}}|\mathbf{H}|_{\hat{g}}\cosh\psi-\alpha_{\hat{e}_{3}}(\nabla\tau)+\psi\Delta \tau\right)\mu_{\Sigma}\\\nonumber 
-\frac{1}{8\pi}\int_{\Sigma}\hat{F}^{a}~_{b}(\hat{e}_{3},\hat{e}_{4})A^{b}~_{a0}\mu_{\Sigma}.
\end{eqnarray}
Luckily, due to the anti-symmetry as a $2-$form on the spacetime, $\hat{F}^{a}~_{b}(e_{3},e_{4})=\hat{F}(\cosh\psi \hat{e}_{3}-\sinh\psi \hat{e}_{4},-\sinh\psi \hat{e}_{3}+\cosh\psi \hat{e}_{4})=\hat{F}^{a}~_{b}(\hat{e}_{3},\hat{e}_{4})$ and therefore is boost invariant. However, notice that the term $\int_{\Sigma}\hat{F}^{a}~_{b}(\hat{e}_{3},\hat{e}_{4})A^{b}~_{a0}\mu_{\Sigma}$ heavily depends on the Yang-Mills gauge choice. Nevertheless, for a stationary space-time, this term is fixed. In addition, this term can be made to vanish if the spacetime does not contain a horizon. This is due to the fact that in the absence of a horizon, $A^{a}~_{b0}$ can always be gauged to zero (such a choice is temporal gauge). However, this term does not vanish for a black-hole spacetime that is electrically charged since a gauge transformation that sets $A^{a}~_{b0}$ to zero would be singular at the horizon. The reference contribution to the gauge part vanishes since $A^{a}~_{b0}$ may be set to zero on a reference spacetime (Minkowski spacetime in this case) which does not contain any horizon.        

\section{Computation of quasi-local energy for Kerr-Newman spacetime} 
In this section, we explicitly compute the Wang-Yau quasi-local energy functional as well as the energy contribution arising from the U(1) gauge sector of the Kerr-Newman spacetime. In Boyer-Lindquist coordinates, the Kerr-Newman metric reads 
\begin{eqnarray}
\label{eq:kerr-newman}
\hat{g}\\
=-\frac{r^{2}-2rM+a^{2}\cos^{2}\theta\nonumber+Q^{2}}{r^{2}+a^{2}\cos^{2}\theta}dt\otimes dt+\frac{r^{2}+a^{2}\cos^{2}\theta}{r^{2}-2rM+a^{2}+Q^{2}}dr\otimes dr\\\nonumber 
+(r^{2}+a^{2}\cos^{2}\theta)d\theta\otimes d\theta+\frac{a\sin^{2}\theta(Q^{2}-2rM)}{r^{2}+a^{2}\cos^{2}\theta}(dt\otimes d\varphi+d\varphi\otimes dt)\\\nonumber 
+\left(\frac{(r^{2}+a^{2})(r^{2}+a^{2}\cos^{2}\theta)+2rMa^{2}\sin^{2}\theta-a^{2}Q^{2}\sin^{2}\theta}{r^{2}+a^{2}\cos^{2}\theta}\right)\sin^{2}\theta d\varphi\otimes d\varphi,
\end{eqnarray}
where $M$, $a$, and $Q$ are the mass, angular momentum, and electric charge, respectively. They are defined in such a way as to have the same dimensions. Let us recognize the important surfaces associated with this spacetime. The horizons are obtained by solving the equation $\frac{1}{\hat{g}_{rr}}=0$ i.e., $r^{2}-2rM+a^{2}+Q^{2}=0$ which yields 
\begin{eqnarray}
R^{+}=M+\sqrt{M^{2}-a^{2}-Q^{2}},~R^{-}=M-\sqrt{M^{2}-a^{2}-Q^{2}}.
\end{eqnarray}
The equation of the ergosphere $\hat{g}_{tt}=0$ yields 
\begin{eqnarray}
R^{+}_{e}(\theta)=M+\sqrt{M^{2}-a^{2}\cos^{2}\theta-Q^{2}},~R^{-}_{e}(\theta)\nonumber=M-\sqrt{M^{2}-a^{2}\cos^{2}\theta-Q^{2}}
\end{eqnarray}
and the ergo region is the region between $R^{+}$ and $R^{+}_{e}$. The outer horizon is the surface $r=R^{+}$. Throughout this article, we will be interested in the region outside of the outer horizon. 

We want to understand the quasi-local energy associated with the constant radius surface $\Sigma$ given by $t=$cosntant,~$r=R$. The induced metric $\sigma$ on $\Sigma$ is explicitly written as follows 
\begin{eqnarray}
\sigma=(R^{2}+a^{2}\cos^{2}\theta)d\theta\otimes d\theta\\\nonumber 
+\left(\frac{(R^{2}+a^{2})(R^{2}+a^{2}\cos^{2}\theta)+2rMa^{2}\sin^{2}\theta-a^{2}Q^{2}\sin^{2}\theta}{R^{2}+a^{2}\cos^{2}\theta}\right)\sin^{2}\theta d\varphi\otimes d\varphi.
\end{eqnarray} 
Fix a basis $(\partial_{\theta},\partial_{\varphi})$ of the tangent space of $\Sigma$ at each point. We will denote the elements of this basis set by $\partial_{a}$ ($a=\theta,\varphi$).  
The corresponding solution of the Maxwell equations yields the connection 
\begin{eqnarray}
A=\frac{Qr}{r^{2}+a^{2}\cos^{2}\theta}dt-\frac{aQr\sin^{2}\theta}{r^{2}+a^{2}\cos^{2}\theta}d\varphi.
\end{eqnarray}
Notice that on this fixed stationary spacetimes, $A_{0}$ is non-vanishing, and therefore the quasi-local energy contribution from the $U(1)$ gauge sector is non-vanishing as well. The curvature $\hat{F}$ of the connection $A$ is found to be 
\begin{eqnarray}
\hat{F}=-\frac{Q(a^{2}\cos^{2}\theta-r^{2})}{(r^{2}+a^{2}\cos^{2}\theta)^{2}}dt\wedge dr-\frac{Qra^{2}\sin2\theta}{(r^{2}+a^{2})^{2}}dt\wedge d\theta\\\nonumber-\frac{Qa\sin^{2}\theta(a^{2}\cos^{2}\theta-r^{2})}{(r^{2}+a^{2}\cos^{2}\theta)^{2}}dr\wedge d\varphi 
-\frac{Qar(r^{2}+a^{2})\sin2\theta}{(r^{2}+a^{2}\cos^{2}\theta)^{2}}d\theta\wedge d\varphi.
\end{eqnarray}
In order to compute the quasi-local energy of a surface $\Sigma$ defined by $t=$constant, $r=R$ that arises from the $U(1)$ sector, we need to fix a pair $(\hat{e}_{3},\hat{e}_{4})$ of normal vectors to $\Sigma$ satisfying $\hat{g}(\hat{e}_{3},\hat{e}_{3})=1,\hat{g}(\hat{e}_{4},\hat{e}_{4})=-1,\hat{g}(\hat{e}_{3},\hat{e}_{4})=0$ 
\begin{eqnarray}
\label{eq:fixed1}
\hat{e}_{3}=\frac{1}{\sqrt{g_{rr}}}\partial_{r},\\
\label{eq:fixed2}
\hat{e}_{4}=\frac{1}{\sqrt{\hat{g}_{\varphi\varphi}(\hat{g}^{2}_{t\varphi}-\hat{g}_{\varphi\varphi}\hat{g}_{tt})}}\left(\hat{g}_{\varphi\varphi}\partial_{t}-\hat{g}_{t\varphi}\partial_{\varphi}\right).
\end{eqnarray}
Clearly they satisfy $\langle \hat{e}_{3},\partial_{a}\rangle_{\hat{g}}=\langle \hat{e}_{4},\partial_{a}\rangle_{\hat{g}}=0$.
An explicit computation yields the following expression for $\hat{F}(\hat{e}_{4},\hat{e}_{3})$ on $\Sigma$
\begin{eqnarray}
\hat{F}(\hat{e}_{4},\hat{e}_{3})|_{\Sigma}=\frac{1}{\sqrt{\hat{g}_{rr}}\sqrt{\hat{g}_{\varphi\varphi}(\hat{g}^{2}_{t\varphi}-\hat{g}_{tt}\hat{g}_{\varphi\varphi})}}\left(\hat{g}_{\varphi\varphi}\hat{F}(\partial_{t},\partial_{r})\nonumber-\hat{g}_{t\varphi}\hat{F}(\partial_{\varphi},\partial_{r})\right)\\\nonumber 
=\frac{1}{\sqrt{\hat{g}_{rr}}\sqrt{\hat{g}_{\varphi\varphi}(\hat{g}^{2}_{t\varphi}-\hat{g}_{tt}\hat{g}_{\varphi\varphi})}}\left(\frac{\hat{g}_{\varphi\varphi}Q(R^{2}-a^{2}\cos^{2}\theta)}{(R^{2}+a^{2}\cos^{2}\theta)^{2}}-\frac{\hat{g}_{t\varphi}Qa\sin^{2}\theta(a^{2}\cos^{2}\theta-R^{2})}{(R^{2}+a^{2}\cos^{2}\theta)^{2}}\right)
\end{eqnarray}
which together with $A_{0}|_{\Sigma}=\frac{QR}{R^{2}+a^{2}\cos^{2}\theta}$ leads to the following expression for the energy density arising from $U(1)$ sector 
\begin{eqnarray}
F(\hat{e}_{4},\hat{e}_{3})A_{0}=-\frac{Q^{2}R}{\sqrt{\hat{g}_{rr}}\sqrt{\hat{g}_{\varphi\varphi}(\hat{g}^{2}_{t\varphi}-\hat{g}_{tt}\hat{g}_{\varphi\varphi})}}\\
\left(-\frac{(R^{2}-a^{2}\cos^{2}\theta)\{(R^{2}+a^{2})(R^{2}+a^{2}\cos^{2}\theta)+(2rM-Q^{2})a^{2}\sin^{2}\theta\}}{(R^{2}+a^{2}\cos^{2}\theta)^{4}}\right.\\\nonumber 
\left.+\frac{a^{2}\sin^{2}\theta(a^{2}\cos^{2}\theta-R^{2})(Q^{2}-2rM)}{(R^{2}+a^{2}\cos^{2}\theta)^{4}}\right)\sin^{2}\theta\\\nonumber 
=\frac{Q^{2}R\sin^{2}\theta}{\sqrt{\hat{g}_{rr}}\sqrt{\hat{g}_{\varphi\varphi}(\hat{g}^{2}_{t\varphi}-\hat{g}_{tt}\hat{g}_{\varphi\varphi})}}\frac{(R^{2}+a^{2})(R^{2}-a^{2}\cos^{2}\theta)}{(R^{2}+a^{2}\cos^{2}\theta)^{3}}.
\end{eqnarray}
We could simply use $\hat{e}_{3}$ and $\hat{e}_{4}$ because $\hat{F}(\hat{e}_{3},\hat{e}_{4})$ is boost invariant as mentioned previously and therefore any orthonormal basis pair $(e_{3},e_{4})$ of the normal bundle would suffice. However, while computing the Wang-Yau quasi-local energy functional that arises from the gravity sector, one needs to make sure that $(\hat{e}_{3},\hat{e}_{4})$ satisfy the desired property (\ref{eq:property}) otherwise we have to choose a different basis that does. 
Collecting all the terms together, we obtain 
\begin{eqnarray}
8\pi \mathcal{QLE}_{gauge}=Q^{2}\int_{\Sigma}\frac{R(R^{2}+a^{2})(R^{2}-a^{2}\cos^{2}\theta)\sin\theta}{(R^{2}+a^{2}\cos^{2}\theta)^{3}}d\theta d\varphi\\\nonumber 
=2\pi Q^{2}R(R^{2}+a^{2})\int_{0}^{\pi}\frac{(R^{2}-a^{2}\cos^{2}\theta)\sin\theta}{(R^{2}+a^{2}\cos^{2}\theta)^{3}}d\theta
\end{eqnarray}
$\mathcal{QLE}_{gauge}$ is manifestly positive and behaves like $\sim \frac{ Q^{2}}{2R}$ as $R\to\infty$. Therefore, one may retrieve the electrical charge as follows 
\begin{eqnarray}
Q:=\lim_{R\to\infty}\sqrt{2R \mathcal{QLE}_{gauge}}.
\end{eqnarray}
If we set $a=0$ (solution reduces to Reissner Nordstrom solution), then the expression for $\mathcal{QLE}_{gauge}$ simplifies tremendously yielding 
\begin{eqnarray}
\mathcal{QLE}_{gauge}=\frac{ Q^{2}}{2R}~\forall R\geq R^{+}.
\end{eqnarray}
We can, in fact, evaluate the integral exactly for any $0\neq a<M$. Direct integration yields 
\begin{eqnarray}
\mathcal{QLE}_{gauge}=\frac{1}{8} Q^{2}R(R^{2}+a^{2})\left(\frac{a^{2}+3R^{2}}{R^{2}(a^{2}+R^{2})^{2}}+\frac{1}{aR^{3}}\tan^{-1}(\frac{a}{R})\right).
\end{eqnarray}
Later we evaluate this term on the outer horizon $R=R^{+}$ assuming a small angular momentum $a$.\\   
Now that we have explicitly computed the contribution of the $U(1)$ sector, we move on to the computation of the quasi-local energy expression for the gravity part. Recall the expression for the quasi-local energy (\ref{eq:gauge-gravity})
\begin{eqnarray}
\label{eq:refphysical}
\label{eq:totalenergy}
\mathcal{QLE}_{gravity}=\frac{1}{8\pi}\underbrace{\int_{\Sigma}\left(\sqrt{1+|\nabla\tau|^{2}_{\sigma}}|\mathbf{H}_{0}|_{\eta}\cosh\psi_{0}-\alpha_{\hat{e}_{30}}(\nabla\tau)\nonumber+\psi_{0}\Delta \tau\right)\mu_{\Sigma}}_{MS}\\-\frac{1}{8\pi}\underbrace{\int_{\Sigma}\left(\sqrt{1+|\nabla\tau|^{2}_{\sigma}}|\mathbf{H}|_{\hat{g}}\cosh\psi-\alpha_{\hat{e}_{3}}(\nabla\tau)+\psi\Delta \tau\right)\mu_{\Sigma}}_{KN}.
\end{eqnarray}
We will explicitly compute each term for the physical spacetime (Kerr-Newman spacetime in the current context) contribution $KN$. For the contribution $MS$ that arises from the Minkowski space, we will use a more direct approach since \cite{Wangyau} derived the following simpler expression for $MS$
\begin{eqnarray}
\label{eq:minkowski}
MS=\frac{1}{8\pi}\int_{\Sigma}\left(\sqrt{1+|\nabla\tau|^{2}_{\sigma}}|\mathbf{H}_{0}|_{\eta}\cosh\psi_{0}-\alpha_{\hat{e}_{30}}(\nabla\tau)\nonumber+\psi_{0}\Delta \tau\right)\mu_{\Sigma}\\ 
=\frac{1}{8\pi}\int_{\hat{\Sigma}}\hat{H}\mu_{\hat{\Sigma}},
\end{eqnarray}
where $\hat{\Sigma}$ is the convex shadow of $\Sigma$ onto the complement of $\partial_{t}$ i.e., on a $\tau=$constant Euclidean slice $\mathbb{R}^{3}$, $\hat{H}$ is the mean curvature of $\hat{\Sigma}$ while realized as a $2-$surface in $\mathbb{R}^{3}$, and $\mu_{\hat{\Sigma}}=\sqrt{1+|\nabla\tau|^{2}_{\sigma}}\mu_{\Sigma}$ is the volume form induced on $\hat{\Sigma}$. Now we proceed to compute the physical space contribution to the Wang-Yau quasi-local energy functional.\\
\textbf{Lemma 2:} \textit{For the Kerr-Newman spacetimes, the basis pair $(\hat{e}_{3},\hat{e}_{4})$ defined in equations (\ref{eq:fixed1}-\ref{eq:fixed2}) is the canonical pair for which $\mathbf{H}=-|\mathbf{H}|_{\hat{g}}\hat{e}_{3}$ and $\langle\mathbf{H},\hat{e}_{4}\rangle_{\hat{g}}=0$}

\textbf{Proof:} Using the definition of $\mathbf{H}$ and the expressions of $(\hat{e}_{3},\hat{e}_{4})$ from (\ref{eq:fixed1}-\ref{eq:fixed2}), we obtain 
\begin{eqnarray}
\langle \mathbf{H},\hat{e}_{4}\rangle_{\hat{g}}=\sigma^{ab}\langle \nabla_{\partial_{a}}\hat{e}_{4},\partial_{b}\rangle_{\hat{g}}=-\sigma^{ab}\langle\nabla_{\partial_{a}}\partial_{b},\hat{e}_{4}\rangle_{\hat{g}}=-\sigma^{ab}\Gamma^{\mu}_{ab}(\hat{e}_{4})_{\mu}\\\nonumber 
=\sigma^{ab}\Gamma^{t}_{ab}\sqrt{\frac{(\hat{g}^{2}_{t\varphi}-\hat{g}_{\varphi\varphi}\hat{g}_{tt})}{{\hat{g}_{\varphi\varphi}}}}
\end{eqnarray}
Explicit calculation shows that $\Gamma^{t}_{\theta\theta}$ and $\Gamma^{t}_{\varphi\varphi}$ vanish for Kerr-Newman spacetime (since $\partial_{t}$ and $\partial_{\varphi}$ are two Killing fields) yielding 
\begin{eqnarray}
\langle \mathbf{H},\hat{e}_{4}\rangle_{\hat{g}}=0.~~~~~~~~~~~~~~~~~~~~~~~~~~~~~~~~~~~~~~~~\Box
\end{eqnarray}
As a consequence of Lemma 2, we do have the necessary ingredients to compute the quasi-local energy corresponding to the gravity sector. The hyperbolic angle $\psi$ is given by 
\begin{eqnarray}
\label{eq:psi}
\psi=\sinh^{-1}(-\frac{\Delta \tau}{|\mathbf{H}|_{\hat{g}}\sqrt{1+|\nabla\tau|^{2}_{\sigma}}}),
\end{eqnarray}
where $|\mathbf{H}|_{\hat{g}}$ is explicitly computed as follows 
\begin{eqnarray}
|\mathbf{H}|_{\hat{g}}=-\langle \mathbf{H},\hat{e}_{3}\rangle_{\hat{g}}=\sigma^{ab}\langle\nabla_{\partial_{a}}\hat{e}_{3},\partial_{b}\rangle=-\sigma^{ab}\langle\hat{e}_{3},\nabla[\hat{g}]_{\partial_{a}},\partial_{b}\rangle_{\hat{g}}\\\nonumber 
=-\frac{\sigma^{ab}}{\sqrt{\hat{g}_{rr}}}\langle\partial_{r},\Gamma^{\mu}_{ab}\partial_{\mu}\rangle_{\hat{g}}=-\frac{\sigma^{ab}}{\sqrt{\hat{g}_{rr}}}\Gamma^{\mu}_{ab}\hat{g}_{\mu r}=-\sqrt{\hat{g}_{rr}}(\Gamma^{r}_{\theta\theta}\sigma^{\theta\theta}+\Gamma^{r}_{\varphi\varphi}\sigma^{\varphi\varphi}),
\end{eqnarray}
where we have used the fact that $\hat{e}_{3}=\frac{1}{\sqrt{\hat{g}_{rr}}}\partial_{r}$ (\ref{eq:fixed1}) and $\langle\hat{e}_{3},\partial_{a}\rangle=0$. Therefore, $|\mathbf{H}|_{\hat{g}}$ may be computed by substituting $\Gamma^{r}_{\varphi\varphi}=(\frac{r}{r^{2}+a^{2}\cos^{2}\theta}+\frac{a^{2}\{r(Q^{2}-rm)+a^{2}m\cos^{2}\theta\}\sin^{2}\theta}{(r^{2}+a^{2}\cos^{2}\theta)^{3}})(r^{2}-2rm+a^{2}+Q^{2})\sin^{2}\theta$ and $\Gamma^{r}_{\theta\theta}=-\frac{r(r^{2}-2rm+a^{2}+Q^{2})}{r^{2}+a^{2}\cos^{2}\theta}$.
Next we want to compute the connection $1-$form of the normal bundle of $\Sigma$ i.e., $\alpha_{\hat{e}_{3}}$.\\ 
\textbf{Lemma 3:} \textit{For an axis-symmetric embedding of a constant radius surface of Kerr-Newman spacetimes into Minkowski spacetime, the contribution of the connection of the normal bundle of $\Sigma$ to the physical energy vanishes i.e., 
\begin{eqnarray}
\alpha_{\hat{e}_{3}}(\nabla\tau)=0.
\end{eqnarray}
}
\textbf{Proof:} 
A simple calculation using the definition of $\alpha_{\hat{e}_{3}}$ yields 
\begin{eqnarray}
\alpha_{\hat{e}_{3}}(\nabla\tau)=\langle\nabla[\hat{g}]_{\nabla\tau}\hat{e}_{3},\hat{e}_{4}\rangle=\langle\nabla_{\sigma^{ab}\partial_{b}\tau\partial_{a}}\hat{e}_{3},\hat{e}_{4}\rangle\\\nonumber 
=\frac{\sigma^{ab}\partial_{b}\tau(\hat{g}_{\varphi\varphi}\Gamma^{\mu}_{ar}\hat{g}_{\mu t}-\hat{g}_{t\varphi}\Gamma^{\mu}_{ar}\hat{g}_{\mu\varphi})}{\sqrt{\hat{g}_{rr}\hat{g}_{\varphi\varphi}(\hat{g}^{2}_{t\varphi}-\hat{g}_{\varphi\varphi}\hat{g}_{tt})}}=-\frac{\sigma^{\varphi\varphi}\partial_{\varphi}\tau \Gamma^{t}_{\varphi r}\sqrt{(\hat{g}^{2}_{t\varphi}-\hat{g}_{\varphi\varphi}\hat{g}_{tt})}}{\sqrt{\hat{g}_{rr}\hat{g}_{\varphi\varphi}}},
\end{eqnarray}
where we have used the fact that $\Gamma^{t}_{\theta r}=0$. 
This term vanishes identically since $\partial_{\varphi}\tau=0$ for an axis-symmetric embedding. ~~~~~~~~~~~~~~~~~$\Box$

In order to compute the corresponding expression for the isometrically embedded surface $\Sigma$ in the Minkowski space, we first need an embedding. Due to the lack of all $4$ degrees of freedom while embedding a $2-$surface isometrically (and therefore 3 constraints) into Minkowski space, the time function $\tau$ is left as a free variable. Therefore, we have to solve for $\tau$ at the end. Wang and Yau \cite{Wangyau} considered the quasi-local mass to be the minimum among all time-like observers (compatible with the notion of rest mass in relativity). This amounts to minimizing the Wang-Yau quasi-local energy in the space of $\tau$, which yields an elliptic equation for $\tau$. As mentioned previously (\ref{eq:minkowski}) Wang-Yau \cite{Wangyau} showed that the reference (Minkowski) contribution (\ref{eq:refphysical})
may be expressed in the following alternative form \begin{eqnarray}
MS=\frac{1}{8\pi}\int_{\hat{\Sigma}}\hat{H}\mu_{\hat{\Sigma}},
\end{eqnarray}
where $\hat{\Sigma}$ is the projection of the embedding on to the $\tau=$constant hypersurface in the Minkowski space and $\hat{H}$ is the total mean curvature of $\hat{\Sigma}$ in $\mathbb{R}^{3}$. The metric $\hat{\sigma}$ of this projected surface reads 
\begin{eqnarray}
\hat{\sigma}=\sigma+\nabla\tau\otimes \nabla\tau.
\end{eqnarray}
Now we consider the isometric embedding of $\Sigma$ into the Minkowski space and compute the embedding functions. assuming axis-symmetry, the embedding may be written as follows 
\begin{eqnarray}
X^{0}=\tau(\theta),X^{1}=A(\theta)\cos\varphi,X^{2}=A(\theta)\sin\varphi,X^{3}=B(\theta),
\end{eqnarray}
which through the isometric condition $\sigma_{ab}=\langle\partial_{a}X,\partial_{b}X\rangle_{\eta}$ yields the following set of ODE 
\begin{eqnarray}
[A(\theta)^{'}]^{2}+[B(\theta)^{'}]^{2}=(\partial_{\theta}\tau)^{2}+\sigma_{\theta\theta},\\
A(\theta)^{2}=\sigma_{\varphi\varphi}.
\end{eqnarray}
In terms of the embedding variables $A(\theta)$ and $B(\theta)$, $\hat{H}$ may be computed explicitly. Through an explicit computation, the complete expression for the quasi-local energy reads 
\begin{eqnarray}
\mathcal{QLE}:=\mathcal{QLE}_{gravity}+\mathcal{QLE}_{gauge}\\
=\frac{1}{4}\int_{0}^{\pi}\left(\frac{2[(\partial_{\theta}\tau)^{2}+\sigma_{\theta\theta}-\frac{(\sigma^{'}_{\varphi\varphi})^{2}}{4\sigma_{\varphi\varphi}}]^{2}+\frac{\sigma^{'}_{\varphi\varphi}}{2}(2\partial_{\theta}\tau \partial^{2}_{\theta}\tau+\sigma^{'}_{\theta\theta}-\frac{[2\sigma_{\varphi\varphi}\sigma^{'}_{\varphi\varphi}\sigma^{''}_{\varphi\varphi}\nonumber-\sigma^{'3}_{\varphi\varphi}]}{4\sigma^{2}_{\varphi\varphi}})}{2[(\partial_{\theta}\tau)^{2}+\sigma_{\theta\theta}][(\partial_{\theta}\tau)^{2}+\sigma_{\theta\theta}-\frac{\sigma^{'2}_{\varphi\varphi}}{4\sigma_{\varphi\varphi}}]^{1/2}}\right.\\\nonumber 
\left.+\frac{[(\partial_{\theta}\tau)^{2}+\sigma_{\theta\theta}-\frac{(\sigma^{'}_{\varphi\varphi})^{2}}{4\sigma_{\varphi\varphi}}]^{1/2}[\sigma^{'2}_{\varphi\varphi}-\sigma_{\varphi\varphi}\sigma^{''}_{\varphi\varphi}]}{2\sigma_{\varphi\varphi}([(\partial_{\theta}\tau)^{2}+\sigma_{\theta\theta})]}\right)d\theta \\\nonumber 
-KN 
+\frac{1}{4} Q^{2}R(R^{2}+a^{2})\int_{0}^{\pi}\frac{(R^{2}-a^{2}\cos^{2}\theta)\sin\theta}{(R^{2}+a^{2}\cos^{2}\theta)^{3}}d\theta,
\end{eqnarray}
where $KN$ is the contribution of the physical spacetime in equation (\ref{eq:totalenergy}) with each term evaluated in terms of $\tau$ and $\sigma$.
It is impossible to exactly solve this integral and obtain a closed-form solution for the total quasi-local energy. However, one may evaluate this on special surfaces such as the outer event horizon ($R=R^{+}$) and on the sphere at spatial infinity with some assumption on the charge and mass of the black hole. However, before moving to explicit asymptotic expansion,  we need to solve the isometric embedding equation in order to fix $\tau$, the time function of the embedding. The following lemma states that $\tau=$constant is a solution to the optimal embedding equation.\\
\textbf{Lemma 4:} \textit{$\tau(\theta,\varphi)$=constant is a solution to the optimal isometric embedding equation (\ref{eq:optimal}).}\\
\textbf{Proof:} The optimal isometric embedding equation reads
\begin{eqnarray}
-(\hat{H}\hat{\sigma}^{ab}-\hat{\sigma}^{ac}\hat{\sigma}^{bd}\hat{h}_{cd})\frac{\nabla[\sigma]_{a}\nabla_{b}\tau}{\sqrt{1+|\nabla\tau|^{2}_{\sigma}}}+\nabla[\sigma]^{a}(\frac{\nabla_{a}\tau \cosh\psi}{\sqrt{1+|\nabla\tau|^{2}_{\sigma}}}|\mathbf{H}|_{\hat{g}}\\\nonumber -\nabla_{a}\psi-(\alpha_{\hat{e}_{3}})_{a})=0,
\end{eqnarray}
where $\psi$ is given by the usual expression (\ref{eq:psi})
\begin{eqnarray}
\psi=\sinh^{-1}(-\frac{\Delta \tau}{|\mathbf{H}|_{\hat{g}}\sqrt{1+|\nabla\tau|^{2}_{\sigma}}}).
\end{eqnarray}
It obvious that for $\tau=$constant, all of these terms except the the last term $\nabla[\sigma]^{a}(\alpha_{\hat{e}_{3}})_{a}$ vanishes. Therefore, it is sufficient to show that $\nabla[\sigma]^{a}(\alpha_{\hat{e}_{3}})_{a}=0$ for an axis-symmetric spacetime such as the one in the current context. Using the definition (\ref{eq:connection}) of the connection $1-$form $(\alpha_{\hat{e}_{3}})$ of the normal bundle of $\Sigma$ in the Kerr-Newman spacetime, we may write 
\begin{eqnarray}
(\alpha_{\hat{e}_{3}})_{a}=\langle\nabla[\hat{g}]_{\partial_{a}}\hat{e}_{3},\hat{e}_{4}\rangle_{\hat{g}}=\frac{(\hat{g}_{\varphi\varphi}\Gamma^{\mu}_{ar}\hat{g}_{\mu t}-\hat{g}_{t\varphi}\Gamma^{\mu}_{ar}\hat{g}_{\mu\varphi})}{\sqrt{\hat{g}_{rr}\hat{g}_{\varphi\varphi}(\hat{g}^{2}_{t\varphi}-\hat{g}_{\varphi\varphi}\hat{g}_{tt})}}.
\end{eqnarray}
Component-wise these read 
\begin{eqnarray}
(\alpha_{\hat{e}_{3}})_{\theta}=\frac{\hat{g}_{\varphi\varphi}\Gamma^{t}_{\theta r}\hat{g}_{tt}+\hat{g}_{\varphi\varphi}\Gamma^{\varphi}_{\theta r}\hat{g}_{\varphi t}-\hat{g}_{t\varphi}\Gamma^{t}_{\theta r}\hat{g}_{t\varphi}-\hat{g}_{t\varphi}\Gamma^{\varphi}_{\theta r}\hat{g}_{\varphi\varphi}}{\sqrt{\hat{g}_{rr}\hat{g}_{\varphi\varphi}(\hat{g}^{2}_{t\varphi}-\hat{g}_{\varphi\varphi}\hat{g}_{tt})}},\\
(\alpha_{\hat{e}_{3}})_{\varphi}=\frac{\hat{g}_{\varphi\varphi}\Gamma^{t}_{\varphi r}\hat{g}_{tt}+\hat{g}_{\varphi\varphi}\Gamma^{\varphi}_{\varphi r}\hat{g}_{\varphi t}-\hat{g}_{t\varphi}\Gamma^{t}_{\varphi r}\hat{g}_{t\varphi}-\hat{g}_{t\varphi}\Gamma^{\varphi}_{\varphi r}\hat{g}_{\varphi\varphi}}{\sqrt{\hat{g}_{rr}\hat{g}_{\varphi\varphi}(\hat{g}^{2}_{t\varphi}-\hat{g}_{\varphi\varphi}\hat{g}_{tt})}}.
\end{eqnarray}
Explicit calculations yield $\Gamma^{t}_{\theta r}=\Gamma^{\varphi}_{\theta r}=0$ for Kerr-Newman spacetime yielding 
\begin{eqnarray}
(\alpha_{\hat{e}_{3}})_{\theta}=0.
\end{eqnarray}
Now since Kerr-Newman spacetime is axis-symmetric, we have 
\begin{eqnarray}
(\alpha_{\hat{e}_{3}})_{\varphi}=(\alpha_{\hat{e}_{3}})_{\varphi}(R,\theta;a,M,Q)
\end{eqnarray}
or 
\begin{eqnarray}
\partial_{\varphi}(\alpha_{\hat{e}_{3}})_{\varphi}=0.
\end{eqnarray}
Therefore the last term yields 
\begin{eqnarray}
\nabla[\sigma]^{a}(\alpha_{\hat{e}_{3}})_{a}=\sigma^{ab}\nabla[\sigma]_{a}(\alpha_{\hat{e}_{3}})_{b}=\sigma^{ab}(\partial_{a}(\alpha_{\hat{e}_{3}})_{b}-\Gamma[\sigma]^{c}_{ab}(\alpha_{\hat{e}_{3}})_{c})\\\nonumber 
=\sigma^{\theta\theta}(\partial_{\theta}(\alpha_{\hat{e}_{3}})_{\theta}-\Gamma[\sigma]^{c}_{\theta\theta}(\alpha_{\hat{e}_{3}})_{c})+\sigma^{\varphi\varphi}(\partial_{\varphi}(\alpha_{\hat{e}_{3}})_{\varphi}-\Gamma[\sigma]^{c}_{\varphi\varphi}(\alpha_{\hat{e}_{3}})_{c})
\end{eqnarray}
and since $(\alpha_{\hat{e}_{3}})_{\theta}=0=\partial_{\varphi}(\alpha_{\hat{e}_{e}})_{\varphi}$, we obtain 
\begin{eqnarray}
\nabla[\sigma]^{a}(\alpha_{\hat{e}_{3}})_{a}=-\sigma^{\theta\theta}\Gamma[\sigma]^{\varphi}_{\theta\theta}(\alpha_{\hat{e}_{3}})_{\varphi}-\sigma^{\varphi\varphi}\Gamma[\sigma]^{\varphi}_{\varphi\varphi}(\alpha_{\hat{e}_{3}})_{\varphi}=0
\end{eqnarray}
since $\Gamma[\sigma]^{\varphi}_{\theta\theta}=\Gamma[\sigma]^{\varphi}_{\varphi\varphi}=0$. This concludes the proof of the lemma.  ~~~~~~$\Box$\\

However, this does not imply that $\tau=0$ is either local or a global minimum of the Wang-Yau quasi-local energy functional (see \cite{chen2014minimizing} for a detailed analysis). It was proven in \cite{chen2014minimizing} that an optimal isometric
embedding is locally unique if the quasi-local mass density is point-wise positive (which is the case here since the Maxwell field satisfies the dominant energy condition). At any rate, the energy corresponding to $\tau=$constant \textit{is} a valid quasi-local energy. One may always perturb $\tau$ about a fixed $\tau=$constant to obtain a local minimum for the Wang-Yau quasi-local energy functional. Since we focus on $\tau=$constant solution now, we need to ensure that the compatibility condition 
$K_{\Sigma}+\frac{\det(\nabla[\sigma]_{a}\nabla_{b}\tau)}{1+|\nabla\tau|^{2}_{\sigma}}>0$ which for $\tau=$constant reduces to $K_{\Sigma}>0$ (such that theorem of Pogorelov \cite{pogorelov1952regularity} applies). This is not obvious since the constant radii surfaces in Kerr-Newman spacetime may develop cone singularity at the pole if the black hole starts rotating rapidly. In such a scenario, the embedding of a closed $2-$surface into the $\tau=$constant Euclidean slice may not be possible since the region near the pole develops negative Gauss curvature. We explicitly compute the Gauss curvature of $\Sigma$ to obtain the necessary condition for which such a problem is avoided. We observe that if $a<\frac{R}{\sqrt{3}}$ for any constant radius surface $\Sigma$, then $K_{\Sigma}>0$ for that surface. We are interested in the region $R\geq R^{+}=M+\sqrt{M^{2}-a^{2}-Q^{2}}>M$ for physical sub-extremal black holes. Therefore if we simply choose $a<\frac{M}{\sqrt{3}}$, then $K_{\Sigma}>0$ everywhere allowing an embedding into a $\tau=$constant Euclidean slice $\mathbb{R}^{3}$. This makes the Wang-Yau quasi-local energy for $\tau=$constant well defined.

Of course, as mentioned previously, the Wang-Yau quasi-local energy for $\tau=$constant is impossible to evaluate exactly. However, if we assume the smallness condition $a\ll M$ (which satisfies the compatibility condition $a<\frac{M}{\sqrt{3}}$), then we may expand several entities such as the metric components and their derivatives in powers of $a/R$ ($R\geq R^{+}>M$). On $R=R^{+}$, several metric components and their derivatives may be expressed in powers of $a/R^{+}$ as follows 
\begin{eqnarray}
\label{eq:expansion1}
\sigma_{\varphi\varphi}|_{R=R^{+}}=\frac{(R^{+2}+a^{2})^{2}\sin^{2}\theta}{R^{+2}}(1-\frac{a^{2}}{R^{+2}}\cos^{2}\theta+O(a^{4}/R^{+4})),\\
\frac{\partial\sigma_{\varphi\varphi}}{\partial \theta}|_{R=R^{+}}=\frac{(R^{2}+a^{2})^{3}\sin2\theta}{R^{4}}(1-\frac{2a^{2}\cos^{2}\theta}{R^{+2}}+O(a^{4}/R^{+4})),\\
\label{eq:expansion4}
\frac{\partial^{2}\sigma_{\varphi\varphi}}{\partial\theta^{2}}|_{R=R^{+}}=(R^{+2}+a^{2})^{2}\left(\frac{2a^{2}\sin^{2}2\theta}{R^{+4}}(1-\frac{2a^{2}\cos^{2}\theta}{R^{+2}}+O(a^{4}/R^{+4}))\right.\\\nonumber 
+\frac{2\cos2\theta}{R^{+2}}(1-\frac{a^{2}\cos^{2}\theta}{R^{+2}}+O(a^{4}/R^{+4}))+\frac{\sin^{2}\theta}{R^{+2}}(\frac{2a^{2}\cos2\theta}{R^{+2}}+O(a^{4}/R^{+4})).
\end{eqnarray}
Utilizing this expansion and the equation of the horizon $2R^{+}M-Q^{2}=R^{+2}+a^{2}$ (and therefore replacing $Q$ in favor of $a$, $R^{+}$, and $M$), we evaluate the quasi-local energy expression. For the Schwarzschild black hole, the quasi-local energy is globally minimized by $\tau=$constant and equals $2M$ (where $M$ is the ADM mass) at the horizon. In such a case, it monotonically decays to the usual ADM mass $M$ at infinity. However, in the current context, one would not expect the quasi-local energy to be equal to $2M$ at the horizon (for $\tau=$constant) due to the presence of ergo region (or equivalently due to the non-zero charge and angular momentum of the black hole). From the physical ground, one would expect that at the outer horizon, the gravitational part of the quasi-local energy (i.e., $\mathcal{QLE}_{gravity}$) should equal to twice of the irreducible mass $2M_{irr}=\sqrt{2M^{2}-Q^{2}+2M\sqrt{M^{2}-a^{2}-Q^{2}}}$ \cite{christodoulou1970reversible, christodoulou1971reversible} of the black hole and it should decay to the usual ADM mass $M$ at infinity (note that $2M_{irr}\geq M$ for sub-extremal and extremal black holes). Assuming a small angular momentum approximation, we indeed obtain that such a result holds. For $a\ll M$, utilizing  (\ref{eq:expansion1}-\ref{eq:expansion4}), we obtain the following asymptotic expansion for the quasi-local energies   
\begin{eqnarray}
\mathcal{QLE}_{gravity}|_{R=R^{+}}=\sqrt{R^{+2}+a^{2}}(1+O(a^{4}/R^{+4})),\\
\mathcal{QLE}_{gauge}|_{R+R^{+}}=\frac{Q^{2}(R^{+2}+a^{2})}{2R^{+3}}(1-\frac{4a^{2}}{3R^{+2}}+O(a^{4}/R^{+4)})
\end{eqnarray}
and therefore $\mathcal{QLE}_{gravity}=\sqrt{R^{+2}+a^{2}}$ modulo higher-order terms on the outer horizon. On the outer horizon, $\sqrt{R^{2}+a^{2}}$ is equal to the twice of the irreducible mass $M_{irr}$. Therefore, for a sufficiently small angular momentum $a$, the Wang-Yau quasi-local energy corresponding to $\tau=$constant is approximately equal to twice the irreducible at the outer horizon. At spatial infinity, we recover the usual ADM mass i.e., 
\begin{eqnarray}
\mathcal{QLE}=M~as~R\to\infty.
\end{eqnarray}
Since $2M_{irr}\geq M$, a natural expectation would be that the total energy exhibits monotonic decay for sub-extremal Kerr-Newman black holes at least in the small angular momentum approximation. This is indeed the case. The $\mathcal{QLE}_{gauge}$ satisfies
\begin{eqnarray}
\partial_{R}\mathcal{QLE}_{gauge}(R)<0,~R\in (R^{+},\infty)
\end{eqnarray}
for all values of angular momentum $a<M$. Since it is in general difficult to evaluate the gravity contribution of the quasi-local energy functional (even for trivial embedding function $\tau=$constant as we have seen previously), we make the approximation $a\ll M$. Using the expansions of type (\ref{eq:expansion1}), we explicitly evaluate $\partial_{R}\mathcal{QLE}_{gravity}$ to yield 
\begin{eqnarray}
\partial_{R}\mathcal{QLE}_{gravity}(R)<0,~R\in (R^{+},\infty).
\end{eqnarray}
From a physical ground, one obvious obstruction to the monotonicity of the Wang-Yau quasi-local energy is the negativity of the gravitational binding energy (true for a general spacetime) and the presence of an ergo region. However, since at the outer horizon (which already lies within the ergo-sphere), the Wang-Yau quasi-local energy already accounts for the possible energy loss at the ergo region by assuming a value of $2M_{irr}$ instead of $2M$ (which would be the case for Schwarzschild black hole where an ergo region is absent). In other words, the Wang-Yau quasi-local energy essentially encodes the information of the energy that can not be extracted from a Kerr-Newman (or Kerr) black hole via Penrose type process. Once this negative energy contribution is accounted for, it is natural to expect that the Wang-Yau quasi-local energy should exhibit a monotonic decay and we confirm such a notion. 

Now if we specialize to the case of zero angular momentum i.e., spherically symmetric static solution (Reissner Nordstrom), then $\tau=$constant is a global minimizer of the Wang-Yau quasi-local energy and in such case, the total energy can be explicitly evaluated and expressed as 
\begin{eqnarray}
\mathcal{QLE}|_{a=0}=R\left(1-\sqrt{1-\frac{2M}{R}+\frac{Q^{2}}{R^{2}}}\right)+\frac{Q^{2}}{2R}\\\nonumber 
\forall R\geq R^{+}= M+\sqrt{M^{2}-Q^{2}}.
\end{eqnarray}
Since $R\left(1-\sqrt{1-\frac{2M}{R}+\frac{Q^{2}}{R^{2}}}\right)>0$, 
\begin{eqnarray}
\mathcal{QLE}|_{a=0}>\frac{Q^{2}}{2R}
\end{eqnarray}
for all values of $R$ on or outside the horizon. Therefore, In the limit of zero angular momentum, we recover the exact form of a weak Bekenstein's inequality. One may compute explicitly on the horizon
\begin{eqnarray}
\mathcal{QLE}(R^{+})|_{a=0}=M+\sqrt{R^{2}-Q^{2}}+\frac{Q^{2}}{2R}\geq Q+\frac{Q^{2}}{2R}
\end{eqnarray}
since $Q\leq M$ for \textit{physical} black holes. Therefore, the equality $\mathcal{QLE}|_{a=0}=\frac{Q^{2}}{2R}$ is never attained. The results obtained so far yield the following theorem regarding the energy content of a Kerr-Newman black hole.

\textbf{Theorem:} \textit{Let $\Sigma$ be a surface of constant radius $R$ in the Ker-Newman spacetime (\ref{eq:kerr-newman}) such that $R\geq R^{+}=M+\sqrt{M^{2}-a^{2}-Q^{2}}, ~a^{2}+Q^{2}\leq M^{2}$. Then the total quasi-local energy associated with the membrane $\Sigma$ satisfies the following strict inequality for all $R\in [R^{+},\infty)$
\begin{eqnarray}
\mathcal{QLE}(R)>\frac{1}{8} Q^{2}R(R^{2}+a^{2})\left(\frac{a^{2}+3R^{2}}{R^{2}(a^{2}+R^{2})^{2}}+\frac{1}{aR^{3}}\tan^{-1}(\frac{a}{R})\right),
\end{eqnarray}
which for large $R$ reduces to 
\begin{eqnarray}
\mathcal{QLE}(R)>\frac{Q^{2}}{2R}+a^{2}Q^{2}O(R^{-3}).
\end{eqnarray}
Moreover, the residual embedding parameter $\tau=$constant satisfies the optimal isometric embedding equation (\ref{eq:optimal}). In such case, the total quasi-local energy ($\mathcal{QLE}$) is expressible in the following form at the outer horizon $R=R^{+}$
\begin{eqnarray}
\mathcal{QLE}(R^{+})=2M_{irr}(1+O(a^{4}/R^{+4}))+\frac{Q^{2}(R^{+2}+a^{2})}{2R^{+3}}(1-\frac{4a^{2}}{3R^{+2}}\\\nonumber+O(a^{4}/R^{+4)})~~~~~for~~a\ll M,
\end{eqnarray}
and 
\begin{eqnarray}
\lim_{R\to\infty}\mathcal{QLE}=M~~~~for~all~a<\frac{M}{\sqrt{3}}.
\end{eqnarray}
In particular, $\frac{\partial\mathcal{QLE}}{\partial R}<0~$ for all $R\in (R^{+},\infty)$ and $a\ll M$. Here $M_{irr}=\frac{1}{2}\sqrt{2M^{2}-Q^{2}+2M\sqrt{M^{2}-a^{2}-Q^{2}}}$ is the irreducible mass of the black hole. In addition, if the angular momentum  vanishes then $\tau=$constant globally minimizes the total quasi-local energy and in such case, it is exactly evaluated as  
\begin{eqnarray}
\mathcal{QLE}|_{a=0}=R\left(1-\sqrt{1-\frac{2M}{R}+\frac{Q^{2}}{R^{2}}}\right)\nonumber+\frac{Q^{2}}{2R},~R\geq R^{+}= M+\sqrt{M^{2}-Q^{2}}.
\end{eqnarray}
}

\section{Concluding Remarks}
Here we have explicitly obtained an expression of the Wang-Yau quasi-local energy for constant radii surfaces in Kerr-Newman black holes. As we have mentioned previously, obtaining a closed-form expression for the Wang-Yau quasi-local mass for a general constant radial surface is almost impossible due to the presence of the embedding function $\tau$ that is to be obtained through solving an elliptic equation the so-called optimal embedding equation. In this particular case, if one assumes an axis-symmetric embedding, one can obtain $\tau=$constant as a solution to the optimal isometric embedding equation. However, such a solution may not be a global or even a local minimum for the quasi-local energy functional \cite{chen2014minimizing}. However, since $\tau=$constant is rigorously obtained as a solution, the associated energy functional the Liu-Yau mass \cite{liu2003positivity, liu2006positivity} still describes a notion of energy bounded by the constant radius membrane. Since $\tau=$constant essentially implies the isometric embedding into a Euclidean slice of the Minkowski space, one needs to ensure that the Gauss curvature of the $2-$ surface $\Sigma$ is everywhere positive. This is indeed guaranteed for an angular momentum that is bounded from above by a suitable factor of the ADM mass. 

Since we are unable to find an explicit meaningful formula for the quasi-local mass for any constant radius surface, we evaluate it on the event horizon and for a large sphere. The total energy contained within the membrane constitutes the pure gravitational energy (may be described by the time component of the Bel-Robinson tensor), the energy of the gauge field coming from its stress-energy tensor as well as the additional pure gauge contribution. For a charged black hole such as the one in the current context, this additional pure gauge contribution does not vanish. This energy is positive everywhere outside the outer horizon (and on it) of the Kerr-Newman black hole and proportional to the square of the charge of the black hole. This contribution essentially provides evidence towards the validity of a weaker version of the Beckenstein type inequality for the Kerr-Newman family of black holes. Even though such an inequality demands $\mathcal{E}^{2}\geq \frac{Q^{4}}{4R^{2}}+\frac{\mathcal{J}^{2}}{R^{2}}$, where $\mathcal{J}$ is the angular momentum, we are able to deduce a version $\mathcal{QLE}_{total}> \frac{1}{8} Q^{2}R(R^{2}+a^{2})\left(\frac{a^{2}+3R^{2}}{R^{2}(a^{2}+R^{2})^{2}}+\frac{1}{aR^{3}}\tan^{-1}(\frac{a}{R})\right)$ since $\mathcal{QLE}_{gravity}> 0$ for a spacetime satisfying dominant energy condition (which is satisfied in the current context) \cite{Wangyau}. Moreover, on special surface such as the outer event horizon, we obtain improved inequality such as $\mathcal{QLE}_{total}\geq M+\frac{Q^{2}(R^{+2}+a^{2})}{2R^{+3}}(1-\frac{4a^{2}}{3R^{+2}}+O(a^{4}/R^{+4)})$ (assuming $a\ll M$) since for a \textit{physical} black hole, $a^{2}+Q^{2}\leq M^{2}$ and $M\geq 0$ by the positive mass theorem of Schoen and Yau \cite{schoen1979proof, schoen1981proof}. While a weaker version of Bekenstein's inequality is really a non-negative definiteness of the entropy of a physical object, the definition of the energy involved is certainly non-unique. In fact, different notions of energy give rise to different coefficients multiplied by the square of the charge. Therefore, a safe conclusion on the physical basis would be that the total energy dominates the square of the charge multiplied by a suitable positive function (of the size of the object under consideration) of the appropriate dimension. A drawback of our study is that we still need to compute the angular momentum contribution in the inequality. However, since Chen, Wang, and Yau \cite{chen2015conserved} have defined a generalized notion of angular momentum associated with a $2-$ surface enclosing a space-like domain in s physical spacetime, we intend to extend this study in the future by adopting their technique to explicitly compute the angular momentum contribution. We note that recently \cite{alaee2019geometric} proved several weak versions of the Bekenstein type inequalities through studying different notions of quasi-local energy.

Assuming a certain smallness condition on the angular momentum, we obtained an asymptotic expression of the total quasi-local energy on the outer horizon. This total energy remarkably agrees (to leading order) with the twice of the irreducible mass of the black hole ($M_{irr}$ \cite{christodoulou1970reversible}) and the pure gauge contribution arising due to electric charge. This result is promising since the notion of quasi-local energy that we adopt here provides a notion of the \textit{true} energy of the black hole that can not be extracted modulo the pure gauge contribution. At asymptotic space-like infinity, the total energy only recovers the ADM mass $M$ as expected (rigorously proven by \cite{wang2010limit} for asymptotically flat Einsteinian spacetimes). If, for the moment, we compare the current scenario with that of Schwarzschild spacetime, then for the latter the quasi-local energy (the one we adopt here) is $2M$ on the horizon and $M$ as one approaches the spatial infinity and decays monotonically from the event horizon to the space-like infinity. In the current context too, we find that the total quasi-local energy monotonically decays from its value $2M_{irr}$ at the outer horizon to the ADM mass $M$ at space-like infinity (notice that $2M_{irr}>M$ for $a,Q<M$). Unlike the Schwarzschild black hole, it does not decay from $2M$ at infinity rather from $2M_{irr}$ which provides an indication that the adopted quasi-local energy encodes the evolution of the true gravitational energy since the irreducible mass ($M_{irr}$) accounts for the energy that the black-hole may lose due to Penrose process \cite{wald1974energy} in the ergo region. Since $\tau=0$ is a solution to the optimal isometric embedding equation which may be perturbed to a local minimum, it is tempting to conjecture that the Wang-Yau quasi-local mass (that the for $\tau$ that truly minimizes the Wang-Yau energy functional) for the Kerr-Newman spacetime monotonically decays from outer horizon to the spatial infinity as well. A further involved study in this direction is left for the future. 

Another important point we note here is that the additional contribution to the quasi-local energy that arises due to the presence of a gauge field is not a gauge independent entity (in the sense of Yang-Mills gauge). In the current context of a fixed stationary spacetime, this gauge dependent property does not make any difference since the \textit{non-physical} gauge variable $A_{0}$ of the Maxwell theory (U(1) gauge theory) is fixed and the presence of a horizon does not allow one to set it to zero through a gauge transformation. It would be worth studying this pure gauge contribution in greater detail in the future and attempting to define suitable gauge invariant energy functional for general spacetime with sources that are electrically charged.   

\section{Acknowledgement} This work was supported by the Center of Mathematical Sciences and Applications (CMSA) at Harvard University.

\address{$^1$ Department of Mathematics, Harvard University, MA, USA\\
$^{2}$ Center of Mathematical Sciences and Applications, Department of Mathematics, Harvard University,~MA, USA}


\end{document}